\documentstyle[12pt]{article}

\newcommand{\be}{\begin{equation}}
\newcommand{\ee}{\end{equation}}
\newcommand{\bea}{\begin{eqnarray}}
\newcommand{\eea}{\end{eqnarray}}

\newcommand{\I} {{\cal H}}

\newcommand{\M} {{\tilde M}}
\newcommand{\V} {{\cal V}_n}

\begin{document}

\begin{center}
\begin{large}
{\bf Big Bangs and Bounces   \\}
{\bf  on the \\}
{\bf  Brane \\}
\end{large}  
\end{center}
\vspace*{0.50cm}
\begin{center}
{\sl by\\}
\vspace*{1.00cm}
{\bf A.J.M. Medved\\}
\vspace*{1.00cm}
{\sl
Department of Physics and Theoretical Physics Institute\\
University of Alberta\\
Edmonton, Canada T6G-2J1\\
{[e-mail: amedved@phys.ualberta.ca]}}\\
\end{center}
\bigskip\noindent
\begin{center}
\begin{large}
{\bf
ABSTRACT
}
\end{large}
\end{center}
\vspace*{0.50cm}
\par
\noindent

\par
In view of our accelerating universe,
one of the outstanding theoretical issues  
 is the  absence of
a  quantum-gravitational 
description  of de Sitter space.
Although speculative,  an intriguing circumvention
may be  found in the realm of brane-world scenarios;
where the physical universe can be interpreted as a 
non-critical 3-brane moving in a higher-dimensional, static bulk. 
In this paper, we focus on the cosmological
implications of a positively curved brane world evolving in the background of 
a ``topological'' anti-de Sitter black hole (i.e, Schwarzschild-like
but with an arbitrary horizon topology).  We show  that
the bulk black hole will typically  induce either an
asymptotically de Sitter ``bounce'' universe or a big bang/big crunch
FRW universe, depending on a critical value
of mass. Interestingly, the critical mass is only non-vanishing
in the case of a spherical horizon geometry.
We go on to provide a  holographic
interpretation of this  curiosity.

\newpage

\section{Introduction}

 Empirical evidence, as deduced from  astronomical data,
 has indicated   
that the physical universe 
is presently  accelerating \cite{bops}. This observation, in turn,
implies that  the universe has a positive
cosmological constant  or, at least,  some exotic form of matter
that mimics this constant in the current epoch.\footnote{An
example of the latter is  a dilatonic scalar field that is slowly
rolling in a stable  potential;
also known as ``quintessence'' \cite{q1,q2}.}
Naturally, one would want  to incorporate this  observational
evidence
 into  our current  understanding   (albeit limited)
 of  quantum gravity and cosmology.
Alas, any such  prospect has proven to be a  formidable challenge 
that is far from being resolved.  Along with the 
roadblock   of explaining the  observed  value of the cosmological
constant, there is an apparent incompatibility
between  a  positive-valued  constant  and 
quantum gravity as we  comprehend  it.
Let us briefly  elaborate on these points.
\\
\par
 {\it Explaining the Cosmological Constant}:  
On a semi-classical level, 
 the cosmological constant is expected to  represent the
vacuum energy density of spacetime. 
Unfortunately,  such expectations fail to
even remotely predict the   observed value of the constant:
$\sim 10^{-120}$ (in units of Planck mass
to the fourth power).
Indeed, this  value is many orders of magnitude smaller than
that of any naive theoretical prediction; 
the most optimistic  being $\sim 10^{-60}$
(if supersymmetry is broken just above the  current accelerator limits).
This discrepancy implies  some type of
subtle  mechanism (or combination thereof) 
that  essentially  cancels the  vacuum density while,
at the same time, leaving  behind a small, very stable residue. 
In spite of many interesting attempts in the literature,
this problem has yet to be  resolved in a way that
does not resort to fine-tuning methods or anthropic 
principles.  For a  discussion on these attempts and the
topic in general,
see \cite{wein,carr} and the references  within.
\par
On a more fundamental level, one might hope that string (or M) 
theory\footnote{For
a  lay-physicist's review on string/M theory and supersymmetry, 
see \cite{polxxx}.}
could, after  a suitable process of  compactification and 
 supersymmetry  breaking, give rise to a stable
vacuum with a positive cosmological constant.
Alas, this does not appear to be the case.  Firstly,
perturbative string theory, with broken supersymmetry,
has inherent  tachyonic 
instabilities. These instabilities, in turn,  
 induce either a  large, negative vacuum energy or 
the decoupling of gravity \cite{banks}. Secondly, for non-perturbative string
theory (or matrix theory \cite{matrix}), the 
vacuum states of interest are plagued by singular behavior;
with this breakdown having been 
 attributed to a relevant ``no-go'' theorem \cite{nunmal}.
Moreover, serious  attempts at circumventing this theorem  
are found to be pathological; 
for instance,  wrong-sign kinetic terms
and non-compact ``compactification'' manifolds. (See \cite{dds}
for further discussion and references.)
\\
\par
{\it  Incompatibility with Quantum Gravity}: Significant
to this discussion is that the current acceleration of
the universe implies an asymptotically (As) de Sitter (dS) 
future.\footnote{Strictly speaking, an AsdS future need not be  the case if 
the acceleration
is   explained by, for instance, a quintessence
theory \cite{q1,q2}. Nevertheless, the same
problematic features, as  discussed in this section, have been shown to persist
in this type of alternative scenario \cite{hksx,fkmpx}.}   
Unfortunately,  many of the quantum-gravitational
aspects of dS spacetimes are poorly understood \cite{banks,witxx,bfxx}.
Much  of the difficulty can be traced to dS space
having a finite value for its entropy \cite{gh2}.  Moreover, this
finite  value
serves as an    upper bound on the 
observable entropy of any  AsdS
spacetime \cite{bou}. Consequently,
it can be  expected that, given a positive cosmological constant,
gravity  will  be described by a finite-dimensional Hilbert space.
This finite size  is, however, directly incompatible
with the infinite-dimensional Hilbert space that is inherent to string theory.
Furthermore,  the same incompatibility issue  arises
when dS  space is considered from a holographic
  perspective. (Note that the holographic principle
\cite{tho,sus} is expected to play a  fundamental role in  
any quantum theory of gravity.)  That is to say,  it has been conjectured
that,
in analogy to anti-dS holography \cite{mal,gub,wit}, 
dS space is dual to a  conformal field theory (CFT).\footnote{In
particular, a Euclidean CFT that lives  on the boundary at  temporal
infinity in dS space  
\cite{str}. There has, since \cite{str},  been a plethora
of  investigations into  this dS/CFT duality. Some very recent 
studies include 
\cite{no222,blah,medyyy,sioxxx,gkvxxx,halyoxxx,kvxxx,cgxxx,awlxxx,no333}. 
Many earlier papers have been referenced
in \cite{med2,medyyy}.} However, any CFT
 should be described (like string theory) by a
Hilbert space of infinite dimensionality.
\par
A further complication that arises in  AsdS
 spacetimes (or any spacetime that gives
rise to a future event horizon \cite{hksx,fkmpx}) is
the unclear status of  physical observables. 
First note that physical observables can  normally\footnote{In this
context, the ``norm'' refers to  asymptotically flat or anti-dS 
spacetimes  (which have a  vanishing or negative 
cosmological constant).}
be defined in terms of S-matrix elements between asymptotically
free particle states \cite{susss,gidd}.
Significant to this picture is the existence of an infinitely 
large 
boundary at spatial infinity. Conversely, AsdS
 spacetimes have no such timelike boundary  and, therefore,
provide no obvious  means for defining  physical observables.
One hypothetical circumvention  is to consider the spacelike  boundary
of an AsdS spacetime
at future infinity. However, because of the inevitable  event
horizon,  this boundary ultimately shrinks to a singular point
from the  perspective  of any given observer \cite{hksx}.
Another possibility  might be to define the S-matrix
as a correlation between states at past and future infinity.
However, such a construction  would require a global view
of the spacetime that is outside the realm of  ``mere
mortals''  \cite{witxx,bfxx}. \\
\par
Although the above discussion paints a bleak picture 
(at least, it was supposed to),
some resolution may be possible in the realm of brane-world scenarios
\cite{ruby}. That is to say, one can consider
the speculative viewpoint that the physical universe is 
``trapped'' on a 3-brane (or 3+1-dimensional hypersurface)
 that
is immersed within  a higher-dimensional   bulk 
spacetime.\footnote{It is
typically  assumed that, of 
the standard model particles,
 only the graviton is allowed to propagate off the brane.  It
is also commonly  assumed that all but one of the
``extra'' bulk dimensions have been compactified to string-scale
length. We will, henceforth, adopt these conventions.}
To be specific,  we will focus on Kraus'  generalization \cite{kra} of
the  Randall-Sundrum model \cite{rs};  whereby
a non-critical   brane is
moving through
 a static,  anti-dS black hole background.\footnote{It is
interesting to note that, although the graviton
is free to propagate into the bulk,  gravity
will typically remain localized on the brane for
the model of interest \cite{wow,psnd}.}   
Let us take  note of the following points of  pertinence to this scenario.  
(i) The brane dynamics will determine the cosmological evolution of
the universe; that is, an observer will
interpret the brane motion as a cosmological contraction
or expansion.
(ii) The cosmological constant is now
an effective one that can be expressed strictly
in terms of bulk parameters; namely, the  curvature
radius of the anti-dS bulk and the brane tension \cite{kra}.
\par
The reason that this brane-world picture may represent an ``improvement''
(over more conventional descriptions)
is that the physical universe  can now, in some holographic sense 
\cite{hmsxxx,ver,sav}, 
be regarded as an anti-dS
one. 
Significantly, the quantum-gravitational aspects of anti-dS theories 
are much better
understood than their dS counterparts.
  Furthermore, 
the   cosmological constant (as measured on the brane)   can now be viewed as
 an  input parameter \cite{banks} that has its origins in the
bulk theory.  One could then hope that such a brane-world picture
will eventually be realized out of a string-compactification 
scenario; perhaps, along the lines of \cite{hver}.
Meanwhile, the extreme (deranged?) optimist 
may  hope that  string theory can even  fix
the  bulk parameters and, therefore,
vicariously fix the brane cosmology!
\par
In the current paper, we investigate, as discussed above, 
the cosmological implications
of  a  non-critical brane moving in a static, 
anti-dS black hole bulk.
 In particular, we are interested in
    the  influence of the  bulk parameters
on  the cosmology of a brane universe with a
positive  cosmological constant. 
This treatment can be viewed as a generalization
of a recent work by Petkou and Siopsis \cite{pet}.
 These authors, however,  studied an anti-dS Schwarzschild
 bulk, whereas  the current analysis
extends considerations to  so-called ``topological'' anti-dS
black  holes \cite{bir}.  Such a black hole solution
is  still Schwarzschild-like, but
the  horizon topology is  allowed to be, in addition to spherical,
flat or hyperbolic.   We also note that  the  two papers
interpret the results from somewhat different perspectives.
\par
The remainder of the paper is organized as follows.
In Section 2,  after  introducing the (arbitrary-dimensional) 
 action
and bulk  solutions of interest, we consider the equation of
motion for the dynamical brane.
 Following prior  literature
(for instance, \cite{kra}), we express this equation
in a form that mimics the standard Friedmann cosmological
equation.
\par
 In Section 3, with guidance from \cite{pet},
we present solutions  for the induced brane metric as a function
of proper time.  (Here, the focus is on   the physically relevant
case of a 3-brane with a 
positive cosmological constant.)   
The form of solution is shown to vary significantly, 
depending on the values of the relevant bulk 
parameters. 
We then classify these various solutions according to
identifiable cosmologies.
 Particularly of interest,  a  ``bounce'' cosmology (i.e.,  a universe
that is AsdS  at  both  past and  future infinity) is only
possible when the  black hole mass stays below
a critical value. Moreover,  this critical value
is only non-vanishing   when the horizon geometry is spherical. 
\par
 In Section 4,  we 
interpret the intriguing observations of the prior section   
from a  holographic perspective.
Section 5 ends with a summary and some further discussion.

\section{A Brane-World Scenario}

To begin the formal analysis, let us consider a  
Randall-Sundrum  type of brane-world scenario \cite{rs}. More specifically,
we are interested in    
 an $n$+1-dimensional anti-dS bulk spacetime that contains
 an $n$-dimensional brane of positive tension. 
Assuming, for sake of simplicity, that there are  no additional
sources of energy/matter on the brane,
we can express the
 gravitational action as follows:
\bea
{\cal I}= {1\over 16 \pi G_{n+1}}\int_{\cal M} d^{n+1}x \sqrt{-g}
\left(R-2\Lambda_{n+1}\right) &+& {1\over 8\pi G_{n+1}}\int_{\partial{\cal M}}
d^{n}x \sqrt{-h}{\cal K}
\nonumber \\
 &+&{\sigma\over 8\pi G_{n+1}}\int_{\partial{\cal M}}
d^{n}x \sqrt{-h}.
\label{1}
\eea
Here,  $G_{n+1}$ is the $n$+1-dimensional Newtonian constant, ${\cal M}$
signifies the bulk manifold, $\partial{\cal M}$  represents the brane manifold,
$g_{\mu\nu}$ is the bulk metric, $h_{\mu\nu}$ is the induced
metric on the brane,
$\Lambda_{n+1}<0$ is the  bulk cosmological constant,  
${\cal K}$ is the trace of the brane's extrinsic curvature,
and $\sigma$ is the  brane tension. 
Note that the  second integral is  analogous to  the Gibbons-Hawking 
surface term \cite{gh} and  is necessary for
a well-defined variational principle on the spacetime boundary
 (in this case,
the brane). 
\par
For future reference, we  include the following useful identities:
\be
\Lambda_{n+1}= -n(n-1)/2L^2, 
\label{1.1}
\ee
\be
h_{\mu\nu}=g_{\mu\nu}-\eta_{\mu}\eta_{\nu},
\label{1.2}
\ee
\be
{\cal K}_{\alpha\beta}= h^{\mu}_{\alpha}h^{\nu}_{\beta}\nabla_{\mu}
\eta_{\nu},
\label{1.3}
\ee
where $L$ is the curvature radius of the anti-dS  bulk
and $\eta_{\mu}$ is the unit normal vector to the brane.
\par
It can be shown  that
 the bulk equations of motion are solved by
the following (static) black hole spacetimes \cite{bir}:
\be
ds^2_{n+1}=-f(r)dt^2+{1\over f(r)}dr^2+r^2d\Omega^2_{k,n-1},
\label{2}
\ee
where:
\be
f(r)= {r^2\over L^2}+k-{\omega_{n}M\over r^{n-2}},
\label{3}
\ee
\be
\omega_{n}={16\pi G_{n+1}\over (n-1)\V}.
\label{4}
\ee
Here,  $d\Omega^2_{k,n-1}$  represents the line element
of an $(n-1)$-dimensional constant-curvature (Euclidean)
hypersurface,  $\V$ is the
dimensionless volume of this hypersurface, and  $k$ and $M$
are constants of integration.
Without loss of generality, $k$ can  be set equal  to +1, 0 or -1;
describing a horizon geometry that is respectively
spherical (i.e., the  anti-dS Schwarzschild case),
flat or hyperbolic.
Meanwhile, $M$ measures the conserved
mass of the black hole and will be regarded
as a non-negative quantity.\footnote{Technically speaking,
the $k=-1$ hyperbolic solution supports a negative value
for the mass \cite{bir}. However, we will disregard this 
controversial scenario, on the grounds that such
a black hole  is known to induce a negative energy in its holographic
boundary theory \cite{caixxx}. Note that such a negative-energy theory
would likely be a non-unitary one.}
\par
As shown elsewhere (for instance, \cite{kra}),  if one
assumes an evolving brane  and an otherwise static bulk,  the brane dynamics
can be formulated so as to describe a
Friedmann-Robertson-Walker (FRW) universe \cite{frw}.
For sake of completeness, let us now run through this  procedure. 
As an appropriate starting point, we introduce  
 Gaussian normal coordinates
in the proximity of the brane:
\be
ds^2_{n+1}=dz^2+h_{\mu\nu}dx^{\mu}dx^{\nu},
\label{5}
\ee
where 
$z=0$ defines the position of the brane. 
Next, the  brane  coordinates  should be
expressed as functions of the
proper time,  $\tau$,  as  measured by a  brane observer. 
In view of  the generic symmetry of the  bulk  solutions,  one
can  write:
\be
x^{\mu}=\left(t(\tau),r(\tau),\theta_{1},...,\theta_{n-1}\right),
\label{6}
\ee
where $\theta_{i}$ denotes the $\tau$-independent
 coordinates of $d\Omega^2_{k,n-1}$.
\par
At this point, it is useful to consider the velocity vector associated
with a comoving test  particle. That is: 
\be
u^{\mu}\equiv {dx^{\mu}\over d\tau}=\left({\dot t},{\dot r},0,...,0\right),
\label{7}
\ee
where here (and throughout) a dot denotes
differentiation with respect to $\tau$.
Considering that this vector must satisfy
$u^{\mu}u_{\mu}=-1$, we  are able to obtain 
 the following relation:
\be
f(r){\dot t}^2-{1\over f(r)}{\dot r}^2=1. 
\label{8}
\ee
\par
Substituting the above result  into Eq.(\ref{2}),
we find that:
\be
ds^2_{n}=-d\tau^2+r^2(\tau)d\Omega^2_{n-1},
\label{9}
\ee
which  notably describes a  metric having a FRW  form.
\par
We can proceed further by  incorporating the unit  vector,
$\eta^{\mu}$, which is normal  to the brane.  
The explicit form of this vector  can be  realized by way of  
a pair of relations:
$\eta^{\mu}\eta_{\nu}=1$ and $u^{\mu}\eta_{\mu}=0$. 
These equations are readily solved to yield:
\be
\eta_{\mu}=\left(-{\dot r},{\dot t},0,...,0\right).
\label{10}
\ee
\par
With the above formalism,
we are now suitably positioned to investigate
the  equations of motion for the brane.  First note that
 the Israel jump conditions \cite{isr}
lead, quite generically, to the following tensor equation (for instance,
\cite{kra}): 
\be
{\cal K}_{\mu\nu}=T_{\mu\nu}-{1\over n-1}T^{\rho}_{\rho}h_{\mu\nu},
\label{11}
\ee
where $T_{\mu\nu}$ is the stress-energy tensor of the brane.
For this matter-free  brane model, the stress tensor is simply
$T_{\mu\nu}=-\sigma h_{\mu\nu}$. Hence,  the above condition
conveniently reduces to:
\be
{\cal K}_{\mu\nu}= {\sigma\over n-1}h_{\mu\nu}.
\label{12}
\ee
\par
We can calculate the extrinsic curvature by way of
the prior formal definition (\ref{1.3}). For current purposes,
it is sufficient to consider the 
component ${\cal K}_{\theta\theta}$;
where  $\theta=\theta_{i}$ for any choice of $i$.
In this case:
\be
{\cal K}_{\theta\theta}= 
\nabla_{\theta}\eta_{\theta}
=-\Gamma_{\theta\theta}^\rho
\eta_{\rho}= rf(r){\dot t}.
\label{13}
\ee
Equating this result with $\sigma r^2/(n-1)$, as prescribed
by Eq.(\ref{12}), we have:
\be
{\dot t}={\sigma r \over (n-1) f(r)}.
\label{14}
\ee
\par
Finally, let us substitute
the above outcome into Eq.(\ref{8}).
Also defining
the Hubble  ``constant'' in the usual way
(i.e., $ H \equiv {\dot r}/r$), we obtain 
the following  equation of motion for the brane:
\bea
H^2&=& -{f(r)\over r^2}+{\sigma^2\over (n-1)^2} 
\nonumber \\
&=&-{k\over r^2}+{\omega_{n}M\over r^n}-{1\over L^2}+{\sigma^2
\over (n-1)^2},
\label{15}
\eea
where Eq.(\ref{3}) has  been  incorporated into the lower line.
\par
Significantly, the above expression - which describes the cosmological
evolution of the brane universe - takes on a Friedmann-like form.
We can make  this connection even more explicit
by  introducing an effective cosmological constant
for the brane universe.  The appropriate definition
turns out to be:
\bea
\Lambda_{n}&\equiv&{(n-1)(n-2)\over 2}\left[{\sigma^2\over (n-1)^2}
-{1\over L^2}\right]
\nonumber \\
&=&{(n-2)\over n}\left[{n\over 2(n-1)}\sigma^2-|\Lambda_{n+1}|
\right],
\label{17}
\eea
as this leads to:
\be
H^2= {2\Lambda_{n}\over (n-1)(n-2)}-{k\over r^2}+{\omega_n M\over r^n}.
\label{16}
\ee
This  above form can, in fact, be identified  with
 the $n$-dimensional Friedmann equation  
for radiative matter (since $\rho_{rad}\sim r^{-n}$
is  standard for an FRW universe).
Notably,  an anti-dS black hole  in the bulk
is, indeed, expected to   induce radiative matter 
in its holographic dual  \cite{ver,sav}. Also of interest,
 the horizon geometry of the black hole
determines the topology of the
spatial slicing in the brane universe.
\par
In the ``traditional'' Randall-Sundrum model \cite{rs},
one is supposed to fine tune the brane tension ($\sigma$) 
so that $\Lambda_n$ precisely vanishes (i.e., a critical
brane). However, there is no reason, {\it a priori},
for  such a fine tuning to occur.   In fact,
current observational evidence
is in favor of a non-vanishing, positive cosmological
constant \cite{bops}. On this basis, we will
assume that  $\Lambda_n>0$  is always satisfied  in the analysis to 
follow.\footnote{For other studies on non-critical, dynamical
brane scenarios, see  \cite{wang,noxxx,pet,pad1,youmxxx,medxxx,pad2}.}
Note that this stipulation translates into
a critical  value for the brane tension. 
More specifically,
$ \sigma^2 > \sigma^2_{c}$, where $\sigma_c=(n-1)/L$.
This condition, in turn, leads to an interesting but
very open question:
can string theory (or any other microscopic framework)
be used to constrain the value of the brane tension?

\section{FRW  Cosmology on the Brane}

In this section, we will  endeavor to
solve the cosmological brane  equation (\ref{16}) 
for all acceptable  values of the bulk  integration
constants; that is, $k$ and  $M$.
To be succinct, let us now focus on the
physically most interesting and analytically solvable
 case of $n=4$ (also,  $\Lambda_n >0$). 
  We do, however, expect that the same
qualitative features will persevere for
larger values of $n$ \cite{pet}; although  
analytical solutions may  no longer be explicitly obtainable.
\par
Let us begin here by appropriately re-expressing Eq.(\ref{16}):
\be
{\dot r}^2={\Lambda_{4}\over 3} r^2 -k+{\omega_4 M\over r^2}.
\label{18}
\ee
It proves to be convenient
if we further incorporate
the following definitions:
${\cal H}^2 \equiv \Lambda_{4}/ 3$ and
$x\equiv r^2$. 
The cosmological equation then takes the form:
\be
{{\dot x}^2\over 4}= {\cal H}^2 x^2-kx +\omega_4 M.
\label{19}
\ee
\par
One finds that the solutions of Eq.(\ref{19})
naturally  separate into distinct classes as 
 $k$ changes discretely  {\it and} as the quantity  
$4\omega_4M {\cal H}^2$ varies
in relation   to unity (or zero).  Regarding  the mass dependence, 
we can demonstrate this behavior  
by rewriting Eq.(\ref{19}) as follows:
\be
{\dot x}^2=4{\cal H}^2\left(x-x_+\right)\left(x-x_-\right),
\label{20}
\ee
where:
\be
x_{\pm}={1\over 2{\cal H}^2}\left[k\pm\sqrt{k^2-4\omega_4 M{\cal H}^2}
\right].
\label{21}
\ee
\par
In view of the above consideration, 
let us define the following (dimensionless)
mass parameter:  $\M\equiv 4\omega_4M{\cal H}^2$. 
It then follows that 
the solutions should be categorized according to
$\M=0$, $0<\M<1$, $\M=1$  and $\M>1$, as well 
as $k=+1$, $k=0$ and $k=-1$.  As it turns out,
Eq.(\ref{19}) can  readily be  solved for each of
these scenarios  ($4\times 3=12$ in all),
and we summarize the results - case by case - immediately 
below.\footnote{Note that we have omitted any solution in which
$r^2<0$ for all  values of $\tau$.} 
\\
\\
CASE (1a) $\M=0$ and $k=+1$: 
\be
r^2={1\over \I^2}\cosh^2(\I\tau).
\label{22}
\ee
CASE (1b) $\M=0$ and $k=0$:
\be
r^2={1\over\I^2}\exp(\pm2\I\tau).
\label{23}
\ee
CASE (1c) $\M=0$ and $k=-1$: 
\be
r^2={1\over \I^2}\sinh^2(\I\tau).
\label{24}
\ee
CASE (2a) $0<\M<1$ and $k=+1$; \\ \\
Scenario (I):
\be
r^2={1\over 2\I^2}\left[1+\sqrt{1-\M}\cosh(2\I\tau)\right],
\label{25}
\ee
Scenario (II):
\be
r^2={1\over 2\I^2}\left[1-\sqrt{1-\M}\cosh(2\I\tau)\right].
\label{25.5}
\ee
CASE (2b) $0<\M<1$ and $k=0$: 
\be
r^2= {\sqrt{\M}\over 2 \I^2}\sinh(\pm 2\I\tau).
\label{26}
\ee
CASE (2c) $0<\M<1$ and $k=-1$: 
\be
r^2={1\over 2\I^2}\left[\sqrt{1-\M}\cosh(2\I\tau)-1\right].
\label{27}
\ee
CASE (3a) $\M=1$ and $k=+1$; \\ \\
Scenario (I): 
\be
r^2={1\over 2 \I^2}\left[1+\exp(\pm 2\I\tau)\right],
\label{28}
\ee
Scenario (II):
\be
r^2={1\over 2 \I^2}\left[1-\exp(\pm 2\I\tau)\right].
\label{28.5}
\ee 
CASE (3b) $\M=1$ and $k=0$: 
\be
r^2= {1\over 2 \I^2}\sinh(\pm 2\I\tau).
\label{29}
\ee
CASE (3c) $\M=1$ and $k=-1$:
\be
r^2={1\over 2 \I^2}\left[\exp(\pm 2\I\tau)-1\right].
\label{30}
\ee
CASE (4a) $\M>1$ and $k=+1$:
\be
r^2={1\over 2\I^2}\left[1+\sqrt{\M-1}\sinh(\pm 2\I\tau)\right].
\label{31}
\ee
CASE (4b) $\M>1$ and $k=0$: 
\be
r^2= {\sqrt{\M}\over 2 \I^2}\sinh(\pm 2\I\tau).
\label{32}
\ee
CASE (4c) $\M>1$ and $k=-1$:
\be
r^2={1\over 2\I^2}\left[\sqrt{\M-1}\sinh(\pm 2\I\tau)-1\right].
\label{33}
\ee
\\
\par
In spite of the  wide array of solutions,
a careful inspection reveals that all of the above
can be classified according to one of six possible
cosmologies.  Let us discuss each of these in turn.
\\ \\
(i) The brane universe is a  ``pure'' dS spacetime, 
which occurs for cases 1a, 1b and 1c. 
 This  particular outcome could  have easily been anticipated, inasmuch as
 $\M=M=0$  translates into an  FRW theory with no matter
and  a  positive cosmological constant; cf. Eq.(\ref{18}).
The alert reader may have noticed that the three solutions
only agree asymptotically.
However,  all three still  describe dS space, just with different
choices of spatial slicing.
(For a  discussion on the various ways of foliating dS space
 and the associated global properties, see \cite{lmmx}.)
The topology of the slicing  is, as one might have expected, 
perfectly correlated with
 the horizon topology of the bulk black hole.
\\ \\
(ii) Depending on the choice
of sign, 
 the brane universe either begins
with a
 ``big bang'' or ends with a ``big crunch''; that is, begins or 
ends at a  singular ($r=0$) surface.\footnote{For the cases in question,
it is readily confirmed that
the curvature (of the induced brane metric) does indeed
 diverge  when $r=0$.  
Hence, these are  ``true'' singularities, beyond which the spacetime
should  not be analytically continued.} 
This occurs for cases 2b, 2c, 3b, 3c, 4a, 4b and 4c.
To be more specific about the form
of these solutions, let us first consider the
 big bang cosmological picture.  In this case,
the brane universe  comes into existence  at a time  
that  can be regarded as $\tau=0$ without loss of
generality.\footnote{The
 singular surfaces in question  do not always coincide with
$\tau=0$. However, $r=0$ does always coincide with a finite value of $\tau$.
Hence,  we are free to rescale 
the proper time coordinate 
to agree  with the usual conventions.} After the bang,
the universe  expands monotonically (with time) 
 towards an AsdS future; that is,
 $r$ grows exponentially as  $\tau\rightarrow +\infty$.
Meanwhile, the big crunch picture is simply the time-reversed version
of the preceding description.
\\ \\
(iii) The brane universe is  AsdS in the infinite past  and
approaches  a static universe (i.e., $H={\dot r}/r=0$)
in the infinite future. (The solutions can also describe the time reverse of
this description, depending on the choice of sign.)
 This quite  unorthodox  cosmology (first recognized
in \cite{pet}) 
 only  occurs   for scenario I of case 3a.
Note that such an occurrence is  highly model dependent;
 requiring a spherical horizon geometry 
with $\M$ set {\it precisely} equal to unity.
\\ \\
(iv) The brane universe  interpolates
between a singularity (either a big bang or big crunch) and
an asymptotically  static universe. This somewhat pathological cosmology
only occurs for  
 scenario II of case 3a and (as discussed above)
is highly model dependent.  
\\ \\
(v) The brane universe  is AsdS 
 in  both the infinite past  and the infinite future (but not  pure dS).
This type of cosmology  strictly occurs for scenario I of case 2a.
Note that
such a universe  can be said to ``bounce''    when
it reverses from a contracting phase  to an expanding one  
 at $\tau=0$.  Further note that, although the size
of the universe is minimal at the bounce surface, it still
does not become vanishingly small.
Such a cosmological
model is very interesting, considering  that  bounce universes 
can not be trivially constructed.  That is to say,
adding a finite energy density to an otherwise purely dS spacetime typically
induces a big bang or crunch \cite{medyyy}.
 \\ \\
(vi) The brane universe  begins with a big bang and ends, after
a finite time period,
with a big crunch. This cosmology only occurs
for scenario II of case 2a. 
\\ 
\par
It  may be of relevance that, besides pure dS space (i),
 only the bounce cosmology (v)
 will  eternally avoid
 the interior  of the  black hole horizon.\footnote{Setting
  $f(r)=0$ and $n=4$ in Eq.(\ref{3}), we find the horizon
location ($r=r_H$) to be given by
$r_H^2={L^2 \over 2}\left[\sqrt{k^2+4\omega_4M/L^2 }-k
\right]$.} 
 The big bang cosmologies, on the other hand, actually
start out at the  black hole singularity.
Although this latter scenario appears problematic, it can still be resolved
with an analytical continuation to Euclidean time \cite{sav}.
As discussed in the cited paper, the  Euclidean version of the 
big bang (brane) universe manages to  remain outside of the bulk horizon 
during its entire evolution.
\par
Generally  speaking, it is interesting to note
that the brane cosmology knows very little about the
 black hole mass, except for its relation to
a pair of critical values; namely,  $\M=0$ and $\M=1$. 
In  support of this notion,  we point out  that
the hyperbolic and exponential functions depend
only on  the proper time, the brane tension 
and the bulk cosmological constant.
(The factors in front of these functions are,
of course,  essentially meaningless.)
Conversely, the brane universe  will always
be able to make a clear
 distinction between different choices of
the bulk topological parameter, $k$.

\section{Holographic Interpretation}

 To help make sense of the  prior  analysis,
let us summarize, for  each type of horizon topology, how
the brane cosmology qualitatively varies as a function
of increasing black hole mass.
\par
Firstly, we  consider the case of a spherical
horizon geometry (i.e., $k=+1$), which translates
into  an  anti-dS Schwarzschild black hole.
Starting at $M=0$ and then ``turning on'' the mass,
we initially observe a  bounce cosmology;
that is, an AsdS spacetime in both the infinite past and infinite
future.\footnote{Note that
a pure dS spacetime, which occurs only for $M=0$, is really  
just  a special type 
of bounce cosmology and will be regarded as such
in this section.  It is also worth noting that the slightest
perturbation to the brane world would cause an $M=0$ geometry
to lose whatever ``esteemed'' status it  did have.} 
As $M$ continues to increase, the bounce
cosmology persists
 until a critical
value of  mass, $M=M_c\equiv (4\omega_4\I^2)^{-1}$, is
reached.\footnote{This critical mass value
was first identified in  \cite{pet}.}  Exactly
at this critical point, one finds a somewhat
pathological cosmology: static in either the  infinite
past or future. On the other hand, for any
$M>M_c$, we observe a typical FRW cosmological 
picture;  either a  spacetime that begins with a big bang
and evolves
towards an AsdS future,
or else the time-reversed (big crunch) scenario.
\par
Secondly, let us consider the flat ($k=0$) and hyperbolic ($k=-1$)
horizon geometries;
that is, the topological analogues \cite{bir} of the anti-dS
 Schwarzchild black hole.
In essence, we observe the same type of mass dependence
 as described above, except that
the critical mass value is now a vanishing one ($M_c=0$) and
there is no longer a pathology at $M=M_c$.
Keep in mind that we are interpreting
a pure dS spacetime as a bounce geometry.\footnote{The reader
may have noticed that
$r=0$ at $\tau=0$ for a dS spacetime with hyperbolic
spatial slicing; cf. Eq.(\ref{24}). However,
this surface is only an apparent singularity (i.e., the curvature remains
finite),
beyond which the spacetime can indeed be
analytically continued \cite{lmmx}. 
Also of interest, the solution for  dS space with  flat
spatial slices (cf.  Eq.(\ref{23}))  only
covers one half of the total manifold. One obtains
the other half by reversing the sign in the
exponential.} 
\par
The generic  existence of such a  bulk mass threshold - effectively separating
bounce from big bang (or big crunch) cosmologies on the brane -
can be  viewed as a direct consequence
of the holographic principle.  To substantiate this
claim, let us next consider the following discussion.
\par
We begin here by recalling (from the literature) a pair of 
relevant holographic bounds.
Firstly, Bousso has demonstrated \cite{bou}  that the finite entropy 
of dS space serves as an upper bound 
on the observable entropy of a spacetime with a positive
cosmological constant.\footnote{This bound was rigorously
verified  for, in particular,  spherically symmetric spacetimes
having  ``physically realistic'' matter \cite{bou}.}
  Secondly,
on the basis of the Bousso bound, 
Balasubramanian, de Boer and Minic  have  conjectured \cite{bdm}
 a  mass bound that also applies to
spacetimes with a positive cosmological constant.
 More to the point,  if such a  spacetime
has  a  total  energy\footnote{For the purposes of this discussion, 
any relevant mass or energy 
 should be defined  in analogy to that  
of Brown and York's quasi-localized energy \cite{by,balxxx}.} 
in excess of pure dS space, then the Balasubramanian {\it et al} 
conjecture predicts the existence of a cosmological
singularity; that is, a big bang or big crunch. 
Thus, a singularity-free AsdS spacetime (i.e., a bounce
cosmology) can only be possible when this mass bound
has  {\it not} been  exceeded.
\par
Let us now apply the above concepts  to  our non-critical
brane universe. It is clear from Eq.(\ref{16}) that
 a massive black hole in the bulk   induces
 an  energy density, in the form of
holographic radiative matter,
 on the brane. Moreover, this holographic brane matter 
is directly proportional to the mass of the
black hole.
Since $M=0$ translates into  a pure dS  brane universe
(cf. Eqs.(\ref{22}-\ref{24})), one might naively 
expect that {\it any} $M>0$ would exceed the 
conjectured mass bound \cite{bdm}.
That is to say, any non-vanishing value of $M$ 
could be expected to 
induce a big bang or big crunch
in the brane cosmology. This observation  fits in very nicely
with the $k=0$ and $k=-1$ scenarios, where
the critical mass value is exactly at zero.
However, the situation is not so pleasantly simple
when we consider a spherical ($k=+1$) horizon geometry.
In this case,  
the bounce cosmology  persists 
 for $M>0$, in spite of an apparent violation of  the 
holographic mass bound.
\par
The resolution of the above paradox lies in the precise
nature of Bousso's entropic bound \cite{bou}; again noting
that this is the critical antecedent for the mass bound
in question \cite{bdm}.  Technically speaking,
the holographic entropy bound limits only the ``accessible''
entropy in a relevant spacetime (i.e.,  one with a positive-valued
cosmological constant).
This  qualification on the entropy makes an important distinction:
accessible degrees of freedom are those which can both influence
and be influenced by a given experiment 
and  inaccessible degrees of freedom are those which cannot.\footnote{The
philosophy that underlies this distinction is
the principle of black hole complementarity \cite{bhc}.
That is,  physical  consistency in a black hole
setting  renders causally inaccessible regions as being
operationally meaningless. 
 In this regard,
it is significant that the causal structure of black hole
 and dS spacetimes are quite similar \cite{gh2}.}
 To put it
another way, any given observer  
 is causally limited to a 
specific domain of a relevant spacetime; that is,  the observer's so-called
 ``causal diamond'' \cite{bou}.
Consequently, when an  observer attempts to  measure the
``entropy of the universe'',  the actual  measurement
will  only reflect the contents of her causal diamond
and not, necessarily, the entire spacetime manifold.
(Keep in mind that accessible entropy is an observer-dependent
 property, so that, in spite of appearances,  the entropic upper bound
is {\it not} really a  global concept.)
\par
In view of the above discussion, it follows that
the upper bound on mass should not, strictly speaking, apply to
the global energy of the spacetime.  Rather,
the litmus test for the bound
should be
 the ``accessible energy''
as determined by a given observer.  Hence, a brane
cosmology with $M>0$ (i.e., with a total energy in excess
of   pure dS space)
will {\it not} necessarily induce a singularity. 
That is to say,  if no observer
can capably measure a mass in excess of the conjectured bound, then
the bounce cosmology should  persist.
It is interesting to note that the existence/non-existence of a singularity
is a global property, whereas the holographic
 mass bound is  most accurately viewed as a local concept.
Nonetheless, this is not a contradiction, insofar as
any given observer may or may not be in causal contact with 
the singularity (if there is one).\footnote{Or to put it
another way, if a tree falls in the forest, does it
always make a sound?}
\par
One question (at least) still remains to be answered; namely,
 why is the critical mass value
 finite when $k=+1$ but otherwise zero? This apparent discrepancy
can be qualitatively accounted for as follows.
First of all,  it  is useful to recall that the  black hole
topology effectively determines the foliation of the
 spacetime on the brane. That is, a black hole with a
spherical/flat/hyperbolic horizon  geometry induces
a brane universe that is foliated by spherical/flat/hyperbolic
spatial slices;  cf. Eq.(\ref{16}).
Now focusing on the flat and hyperbolic cases,  let us point out
that the constant-time slices in these spacetimes are
typically  infinite in extent (since
$d\Omega^2_{k\leq 0,n-1}$ contains an unbounded coordinate).
 Hence, even  the slightest 
increase in energy density on the brane (i.e., the
slightest increase in $M$) can lead to significant
perturbations in the underlying  dS geometry.
Hence, on an intuitive basis, 
even  an infinitesimal value for $M$  can be expected
to violate  the  mass bound (for
at least one hypothetical observer) and thus collapse the spacetime.
\par
Next, let us  concentrate on the case of a spherical topology.
The constant time slices on the brane are now strictly compact
(since  $d\Omega^2_{k=1,n-1}$ contains only angular
coordinates). Hence, a relatively small increase
 in  energy density on the brane should
give rise to a  proportionally small increase in an observer's
accessible energy.
 It thus follows, again in an intuitive sense,
that the underlying dS geometry (for $k=1$) should be
somewhat resilient to an increasing 
black hole mass, as long as $M$ does not get too large.  
This notion is, of course,
in agreement with the analysis of the prior section.
To reiterate, the  bounce universe can persist until
reaching some  critical value for the  energy density; 
 above which, the holographic mass bound finally fails
and a singular collapse  ensues.
 It is also worth pointing out that,  for general
FRW cosmologies (i.e., independent of the brane-world scenario), 
bounce universes
are   significantly more   
favorable  when the  spacetime is foliated by
spherical slices \cite{kklt}.  

\section{Conclusion}

In summary, we have considered the  cosmological implications 
of a  dynamical brane-world scenario.  In particular,
we have studied a non-critical brane moving
in the static background of an  anti-de Sitter  black hole.
 The novelty
of our treatment, as compared to an earlier work \cite{pet},
was in allowing for all possible horizon topologies  
 of  the  Schwarzschild-like  black hole in the bulk. 
\par
After introducing the relevant action and bulk solutions (of
arbitrary dimensionality),
we went on to formulate the  equation of motion for
the brane. Moreover,  
 we were able to
recast  this expression  into a form that mimics the  Friedmann equation
for radiative matter. Significantly,
the energy density of this holographically
induced matter is directly proportional
to the mass of the bulk black hole.  
Meanwhile, the horizon geometry of the black hole
fixes  the topology of the spatial slices
in the brane universe.
\par
In the next phase of the analysis, we
focused on the  physically relevant scenario of a 4-dimensional
brane universe with a 
 positive cosmological constant \cite{bops}.
Under these conditions, we were able to
re-express the brane cosmological equation in a readily solvable
form. Various solutions were presented for all relevant values
of the  parameters, $M$ and $k$, describing the bulk geometry.
The most interesting feature was the existence
of a critical
value for the black hole  mass, $M=M_c$,  which is 
 finite  for a spherical ($k=+1$) topology   but
is otherwise ($k=0$, $k=-1$) vanishing.
Significantly, when $M \leq M_c$, one obtains singularity-free, 
 asymptotically de Sitter bounce cosmologies
(including pure de Sitter space when $M=0$).\footnote{One notable
exception being the finely tuned case of $M=M_c$ and $k=+1$.
This choice of bulk parameters yields a somewhat pathological cosmology
that is asymptotically static in either the infinite past
or the future.} 
 Conversely, if
$M > M_c$, then only FRW cosmologies having a big bang
(or big crunch) are possible.  
\par
In the penultimate  section of the paper, 
we went on to interpret this critical mass 
from a holographic perspective. In particular,
we  utilized holographic bounds
that limit the accessible entropy \cite{bou}
and matter \cite{bdm} of any asymptotically de Sitter spacetime.
On this basis, we were able to provide ample justification for  
 the generic existence of such a critical mass value. 
Moreover,  holographic arguments  enabled us
to explain why  $M_c$ takes  on a finite value
exclusively in the $k=+1$  case.
\par
It should be clarified that our holographic interpretation
was conspicuously   of a qualitative nature. 
It would be interesting  if one could  further apply
holographic arguments on a quantitative level
and directly predict
the critical mass values.  
 In this regard, 
we note the existence of two distinct holographic dualities
that could well come into play. These being: (i) a dual relation
between the anti-de Sitter bulk and a conformal
field theory (CFT) living on the brane \cite{mal,gub,wit}
and (ii) a  duality between the de Sitter brane and
a Euclidean CFT that lives at temporal infinity
\cite{str,bdm}.  However,  progress along these lines
may be impeded by a  limited understanding of
the dual boundary theories. That is to say, in spite
of the  success of these holographic dualities,
the  CFTs in question
can best be classified as   abstractions. Moreover, the de Sitter-based
correspondence currently has an unclear status
\cite{dls}.
Nevertheless, we hope to address this  intriguing matter
in a future study.  
\par
On a less speculative note, one might extend the prior
treatment by considering  more exotic  black
holes in the bulk or by  introducing  
other forms of matter onto the brane.
With regard to the latter scenario, possibilities
include conventional dust matter, stiff matter (which also appears as
a natural consequence of electrostatic charge in the bulk  \cite{bmw})
and a $M^2$  term. This last contribution appears
 when    the induced brane energy is directly calculated
via a Hamiltonian method \cite{pad2}. Once again,
we defer such prospects  to a  future time.

\section{Acknowledgments}
\par
The author  would like to thank  V.P.  Frolov  for helpful
conversations.



\end{document}